\documentclass{an}
\usepackage{graphicx}
\usepackage{times}
\usepackage{fancyhdr}
\begin{document}

\title{FIASCO - A new Spectrograph at the University Observatory Jena\thanks{Based on
observations obtained with telescopes of the University Observatory Jena, which is operated by the
Astrophysical Institute of the Friedrich-Schiller-University.}}

\author{Markus~Mugrauer\inst{1} \& G. Avila\inst{2}} \institute{Astrophysikalisches
Institut und Universit\"{a}ts-Sternwarte Jena, Schillerg\"{a}{\ss}chen 2-3, 07745 Jena, Germany \and
European Southern Observatory, Karl-Schwarzschild-Strasse 2, 85748 Garching bei M\"{u}nchen, Germany}

\date{Received; accepted; published online}

\abstract{A new spectrograph (FIASCO) is in operation at the 0.9\,m telescope of the University
Observatory Jena. This article describes the characterization of the instrument and reports its
first astronomical observations, among those Li (6708\AA)~detection in the atmosphere of young
stars, and the simultaneous photometric and spectroscopic monitoring of variable stars.}
\correspondence{markus@astro.uni-jena.de}

\maketitle

\section{Introduction}

The University Observatory Jena is located close to the small village Gro{\ss}schwabhausen, west of the
city of Jena. The Friedrich Schiller University operates there a 0.9\,m reflector telescope which
is installed at a fork mount (see e.g. Pfau 1984)\nocite{pfau1984}. The telescope can be used
either as a Schmidt-Camera, or as Nasmyth telescope. In the Schmidt-mode ($f/D=3$) the telescope
aperture is limited to the aperture of the installed Schmidt-plate $D=0.6$\,m. In the Nasmyth-mode
the full telescope aperture $D=0.9$\,m is used at $f/D=15$.

At the end of May 2008 a new fiber spectrograph with an external calibration unit saw its first
light at the University Observatory Jena. The spectrograph FIASCO (\textbf{\large F}iber
\textbf{\large A}mateur \textbf{\large S}pectrograh \textbf{\large C}asually \textbf{\large
O}rganized) was developed and built by the \textbf{\large C}lub of \textbf{\large A}mateurs in
\textbf{\large O}ptical \textbf{\large S}pectroscopy (CAOS, see e.g. \cite{avila1999}) working at
the European Southern Observatory and the Max Planck Institute of Extraterrestrial Physics in
Garching close to Munich.

In this paper, first we describe the main characteristics of the new spectrograph. In the second
section the internal layout, and all external components of FIASCO are described. We present the
spectroscopic properties of the spectrograph as used at University Observatory Jena, as well as the
characterization of its CCD detector. Finally, in the third section we present first results
obtained during test observations with the instrument after its first light end of May 2008.

\section{FIASCO Characterization}

\subsection{The Spectrograph and its Calibration Unit}

\begin{figure}[h!]
\resizebox{\hsize}{!}{\includegraphics[angle=0]{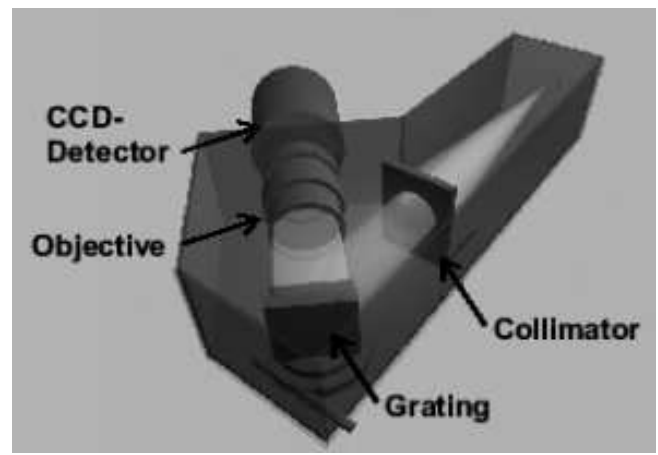}} \caption{A 3d model of FIASCO with
all its internal optical components indicated with black arrows.} \label{fiasco3d}
\end{figure}

The spectrograph is installed in a room directly below the mount of the 0.9\,m telescope and is
operated from the control room of the University Observatory. A 3d model of the spectrograph
internal layout is shown in Fig.\,\ref{fiasco3d}, with all its optical components.

The spectrograph is coupled to the 0.9\,m telescope by an optical fibre. The fibre beam is
collimated by a 50\,mm diameter doublet acting as collimator ($f/D=6$), and the dispersion is
achieved with a reflective diffraction grating with 1300\,lines/mm. Finally, a 200\,mm focal
distance photographic objective open to $f/D=4$ re-image the spectrum on the CCD detector, which is
a SITe TK1024 CCD sensor with 1024$\times$1024 pixel, each a square with an edge length of
24\,$\mu$m. The detector is Peltier cooled, and cooling temperatures 40\,K below ambient air
temperature can be reached.

The optical fibre which connects the spectrograph with the 0.9\,m telescope has a diameter of
70\,$\mu$m, and a length of 15\,m. One end of the fibre is directly connected to the spectrograph
while the other end is coupled to the Nasmyth telescope beam through a mini-lens. This lens
projects the pupil of the telescope on the fibre in such a way to reduce the telescope $f/15$ beam
into a $f/6$ one in the fibre. This optical coupling reduces significantly the so-called focal
ratio degradation by the fibre. In order to put the image of the star in front of the lensed fibre,
a thin 100\,mm diameter nickel reflective plate with a 200\,$\mu m$ hole in the center is placed in
front of the lens. The plate is inclined to send the telescope beam to a fiber-viewing camera.
Figure\,2 shows the fiber coupler, holding the fibre, mini-lens, nickel plate and the fiber-viewing
camera. Between the fiber coupler and the Nasmyth interface flange, a calibration unit is
installed. The unit includes spectral calibration and flatfielding lamps. This calibration unit was
build at the Astrophysical Institute of University Jena, and can be operated from the telescope
control room.

\begin{figure}[h!]
\resizebox{\hsize}{!}{\includegraphics[angle=0]{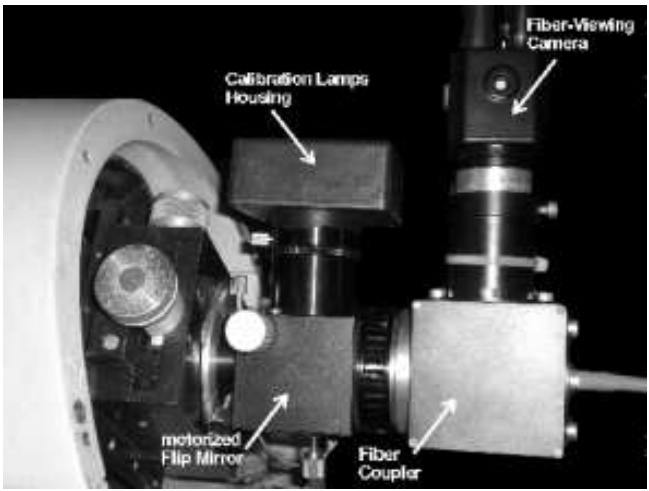}} \caption{The FIASCO calibration unit
and the fiber coupler installed at the Nasmyth interface flange of the 0.9\,m telescope. The
individual components are indicated with white arrows.} \label{calib}
\end{figure}

The fiber-viewing camera is a sensitive video camera, which is equipped with an interline transfer
CCD sensor with 752$\times$582 pixel (8.6\,$\mu$m$\times$8.3\,$\mu$m). The camera redout
electronics offers periodically frame adding from 1 up to 256 frames each with 1/25\,s of
integration time, i.e. integration times of 1/25\,s up to 10.24\,s. With the periodical data
readout stars as faint as $V\sim12$\,mag can be detected with the fibre-viewing camera. In addition
for telescope acquisition of fainter sources also endless frame adding within freely adjustable
time intervals is possible. The signal of the fibre viewing camera is displayed on a monitor in the
telescope control room and shows the Nasmyth focus of the 0.9\,m telescope with the fibre entrance.
This allows an observer, during telescope acquisition, to place a target precisely on the fibre
entrance so that all light, collected by the 0.9\,m telescope, can reach the spectrograph.

The calibration unit is composed of an electrically controlled flip-mirror and the calibration
lamps housing. For calibration the flip-mirror can be moved in the light pass of the telescope. The
mirror then reflects the light of the calibration lamps directly on the fibre entrance. A Ne arc
lamp is used for wavelength calibration and a tungsten lamp for flatfielding, respectively. The
brightness of both lamps was adjusted so that calibration spectra with sufficiently high
signal-to-noise ($S/N>100$) can be achieved already in 10\,s of integration time.

In order to achieve a precise focusing of the 0.9\,m telescope a new focus electronic was developed
and built at University of Jena, which is also operated from the telescope control room. Precise
focusing is of great importance in order to minimize light losses due to unfocused and therefore
broadened telescope PSF. The used 70\,$\mu$m fibre covers a circular area on the sky with a
diameter of 2.7\,arcsec, which is sufficiently large to grab all light collected by the 0.9\,m
telescope under typical seeing conditions at University Observatory Jena.

\subsection{The FIASCO CCD Detector}

\begin{figure}[h]
\resizebox{\hsize}{!}{\includegraphics[]{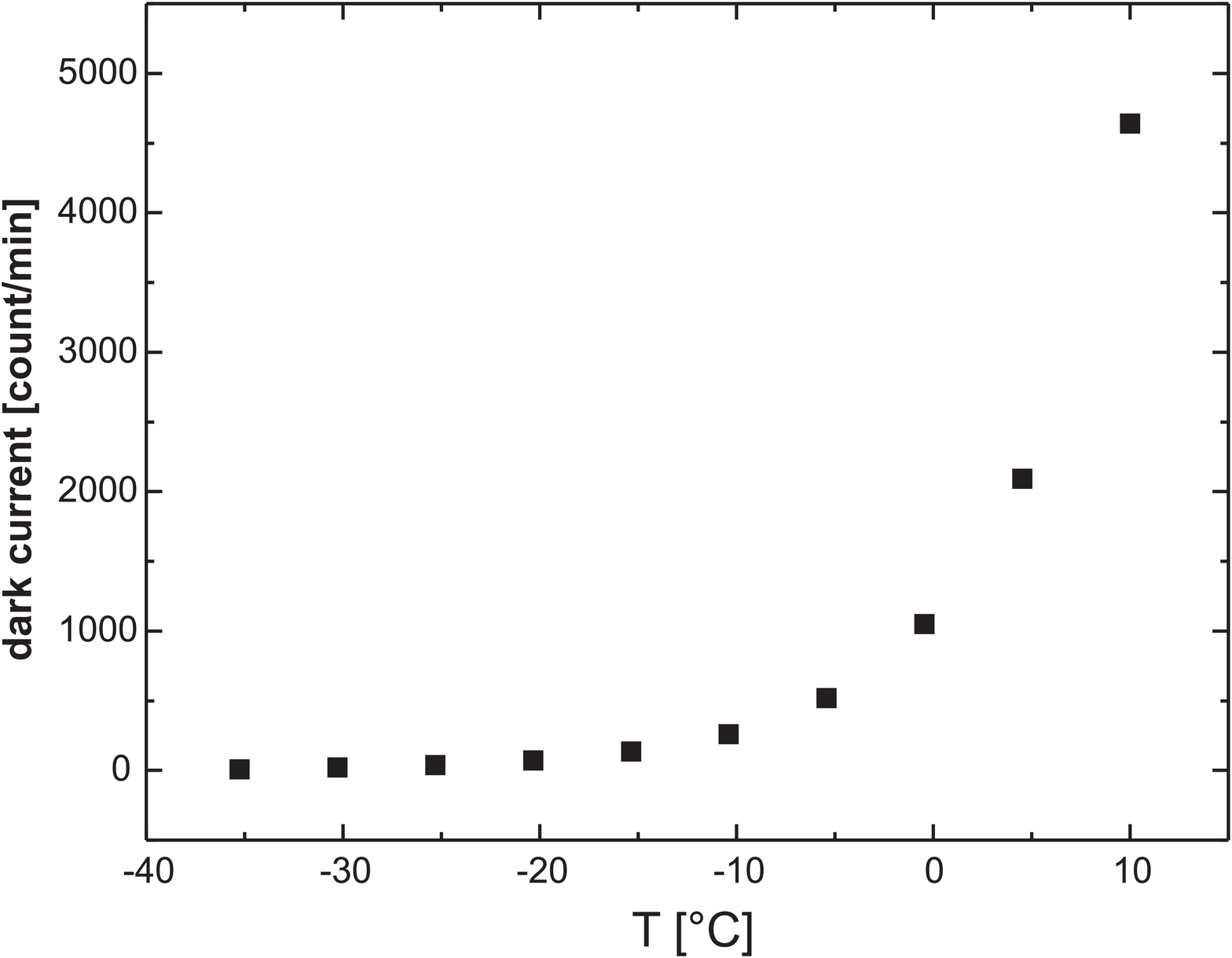}} \caption{The dark current of the FIASCO
detector for a range of detector cooling temperatures. The dark current is exponentially dependant
on the cooling temperature and remains on a low level of less than 100\,count/min below the
critical cooling temperature of $-20\,^{\circ}$C.} \label{dark}
\end{figure}

As described above, the CCD sensor installed in FIASCO is Peltier cooled and the detector cooling
temperature is monitored and regulated by a cooling electronic to stabilize the detector dark
current. For a chosen detector temperature the detector cool down and the temperature stabilization
are completed after 10 minutes. The dark current of the CCD detector, as measured for a range of
cooling temperatures, is illustrated in Fig.\,\ref{dark} and listed in Table\,\ref{darktab}.

\begin{table}[htb]
\centering\caption{The dark current of the FIASCO detector for a range of detector cooling
temperatures. For cooling temperatures below \newline-20\,$^{\circ}$C the FIASCO detector exhibits
a low dark current of less than 100\,count/min.}
\begin{tabular}{c|c|c|c}
\hline
T & dark current & T & dark current\\
$[^{\circ}$C] & [count/min] & $[^{\circ}$C] & [count/min]\\
\hline
$-35$ & $9$   & $-10$   &  $261$\\
$-30$ & $19$  & $-05$   &  $518$\\
$-25$ & $37$  & $~~~00$ & $1046$\\
$-20$ & $72$  & $+05$   & $2093$\\
$-15$ & $136$ & $+10$   & $4642$\\
\hline\hline
\end{tabular}
\label{darktab}
\end{table}

As expected, the dark current is an exponential function of the used detector cooling temperature
and rapidly increases with temperature. We obtain as fit to the measured dark current values at
different cooling temperature:  $log(I[counts])=3.05264 + 0.05925 \cdot T[^{\circ}C]$. While the
dark current is high for detector cooling temperatures over -10\,$^{\circ}$C, where it reaches
already several hundreds of count/min, it remains below 100\,count/min for cooling temperatures
below -20\,$^{\circ}$C. Hence, always cooling temperatures as low as possible but in particular
below this critical temperature value should be chosen. The integrated Peltier cooling reaches this
temperature range even in warm summer nights.

\begin{figure}[h!]
\resizebox{\hsize}{!}{\includegraphics[]{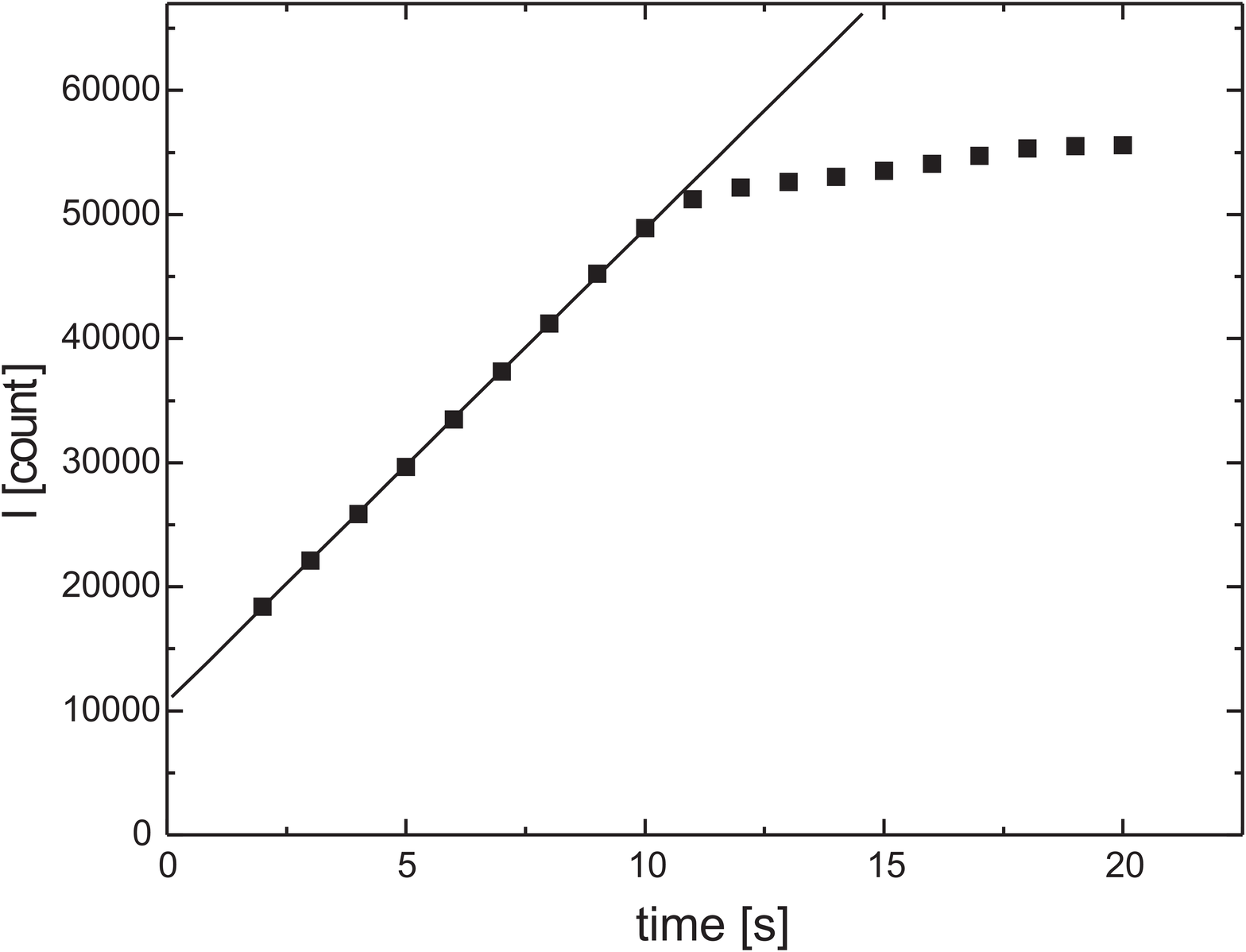}} \caption{The linearity of the FIASCO detector.
The detected signal $I$ of lamp flats is plotted for a range of integration time. The detector
exhibits a high linearity (see straight line) from the bias level at about 11000\,counts, up to a
detector signal of 50000\,count.} \label{lin}
\end{figure}

The detector linearity was measured with flatfield spectra, obtained with the installed calibration
unit. The dependency of the detected signal $I$ of the flatfield spectra on the integration time is
shown in Fig.\,\ref{lin}.

\begin{figure}[h!]
\resizebox{\hsize}{!}{\includegraphics[]{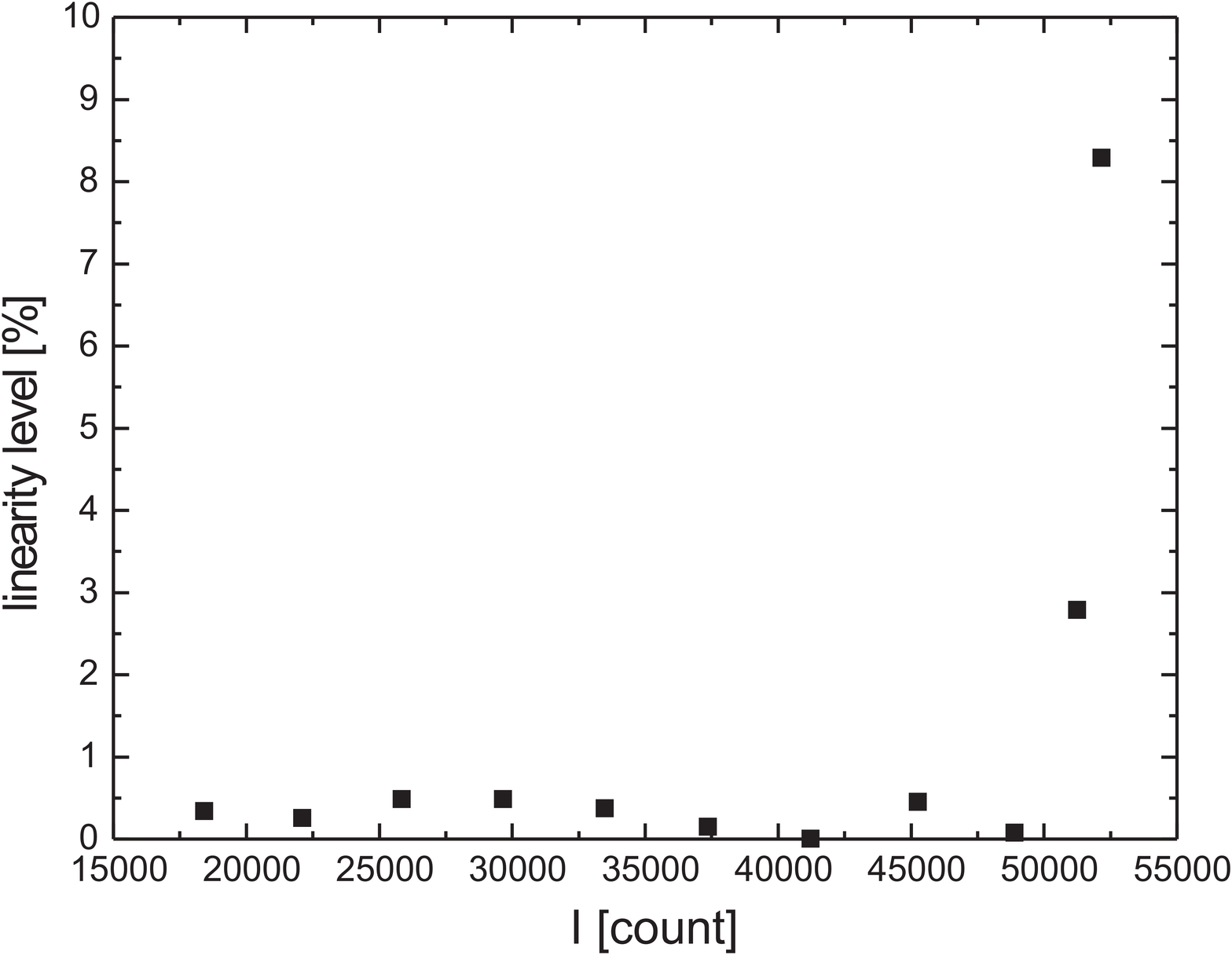}} \caption{The lineartity level of the FIASCO
detector for a range of detector signal $I$.} \label{linlev}
\end{figure}

The measured signal on the detector increases highly linear from the level of the detector bias
around 11000\,count up to 50000\,count. The linearity level, i.e. the deviation in percent from
perfect linearity, is less than 1\,\% in this range of signal (see Fig.\,\ref{linlev}). In
contrast, for signals higher than 50000\,count the detector rapidly looses its linearity. Hence,
the detector signal should always be limited below the critical value of 50000\,count.

\subsection{Spectroscopic properties of FIASCO}

The spectrograph can be used over a wide spectral range limited by the size of the CCD detector. By
rotating the installed reflective grating large shifts in the spectral range can be achieved.
Although most observations are centered around the H$\alpha$-line at 6563\,\AA.

\begin{figure}[h]
\resizebox{\hsize}{!}{\includegraphics[]{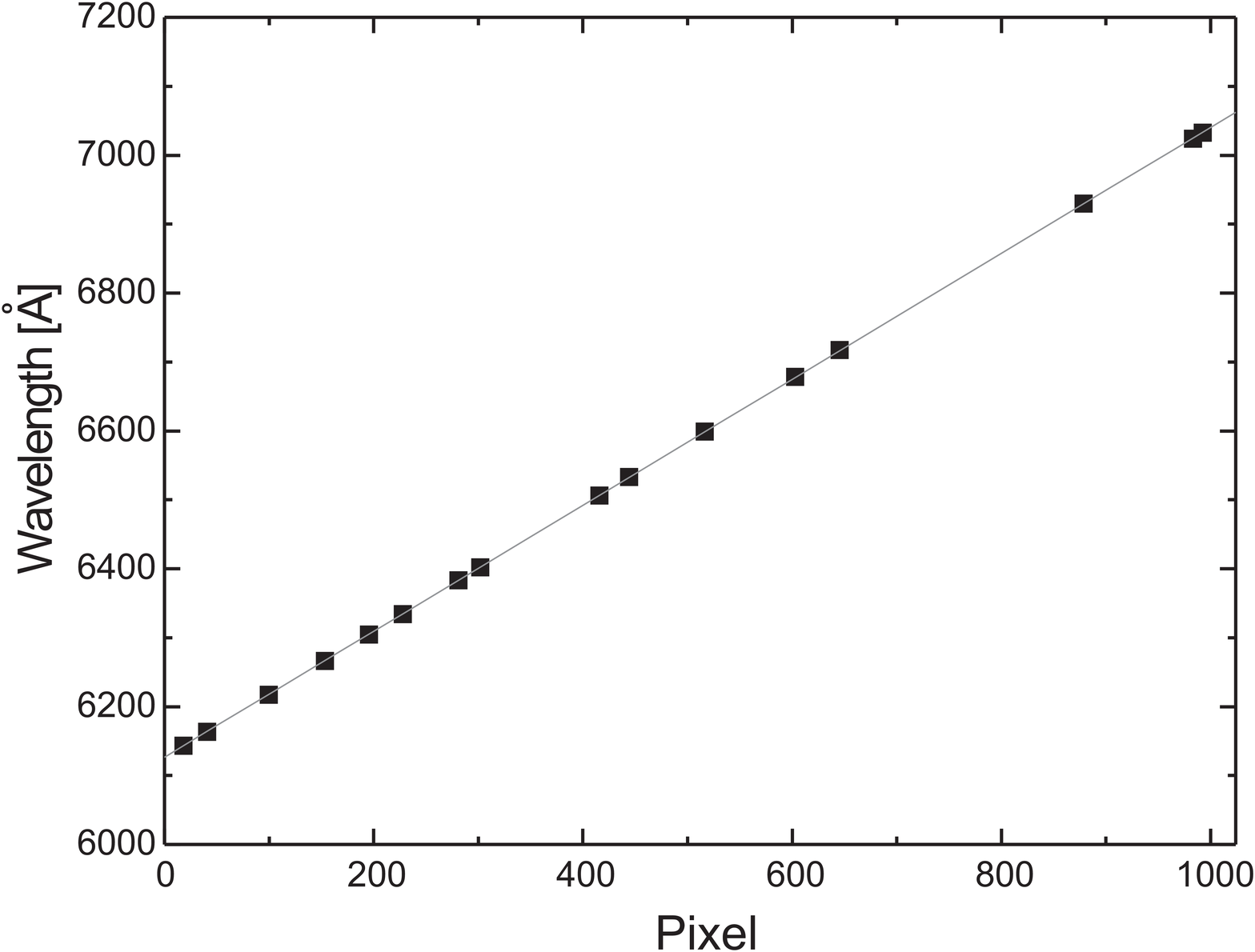}} \caption{The wavelenghth calibration of FIASCO.
The wavelength calibration of FIASCO is a highly linear function over the whole covered spectral
range from 6127\,\AA~to 7061\,\AA.} \label{disp}
\end{figure}

The CCD detector is well aligned with the dispersion direction of the grating. The measured
deviation of the spectrum on the detector perpendicular to the dispersion direction is less than 1
pixel, i.e. 24\,$\mu$m, over the whole detector length (1024 Pixel or 24.6\,mm). In
Fig.\,\ref{disp} we show the wavelength calibration of a typical spectrum obtained with the Ne arc
lamp of the calibration unit.

The dispersion is highly linear over the whole chosen spectral range. We obtain
0.9137$\pm$0.0005\,\AA~per pixel as dispersion over the whole covered wavelength range, which spans
from 6127\,\AA~up to 7061\,\AA, i.e. $\Delta \lambda=934$\,\AA.

\begin{figure}[h]
\resizebox{\hsize}{!}{\includegraphics[]{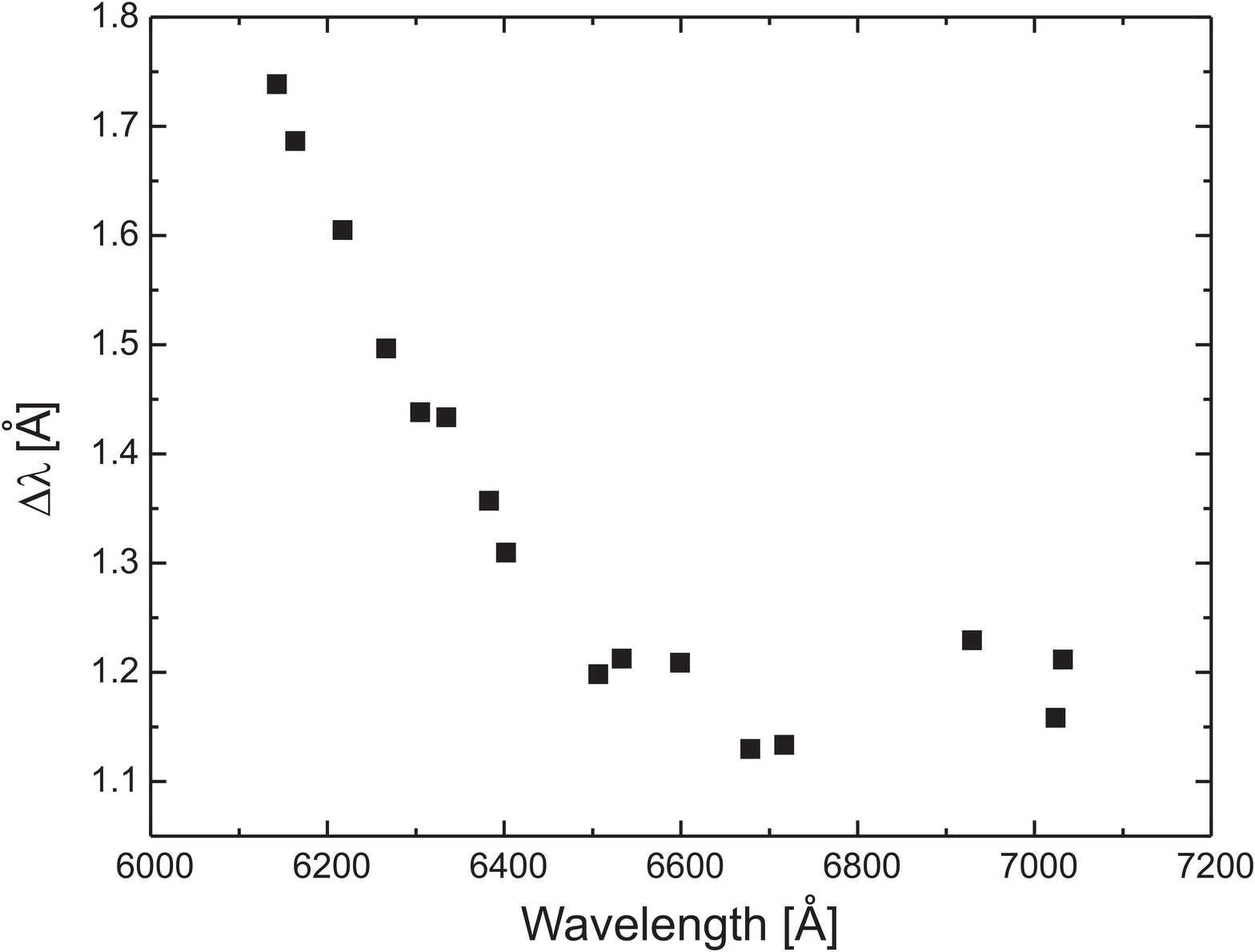}} \caption{The spectral resolution of FIASCO for
the used spectral range between 6127\,\AA~and 7061\,\AA.} \label{res}
\end{figure}

By measuring the full width half maximum of all Ne lines in the wavelength calibration spectrum we
could determine the spectral resolution over the covered spectral range. The measured spectral
resolution for a range of wavelength is illustrated in Fig.\,\ref{res}. The spectral resolution
decrease from about 1.8\,\AA~at 6127\,\AA~to $\sim$1.2\,\AA~for wavelength longer than 6500\,\AA.

\begin{figure}[h]
\resizebox{\hsize}{!}{\includegraphics[]{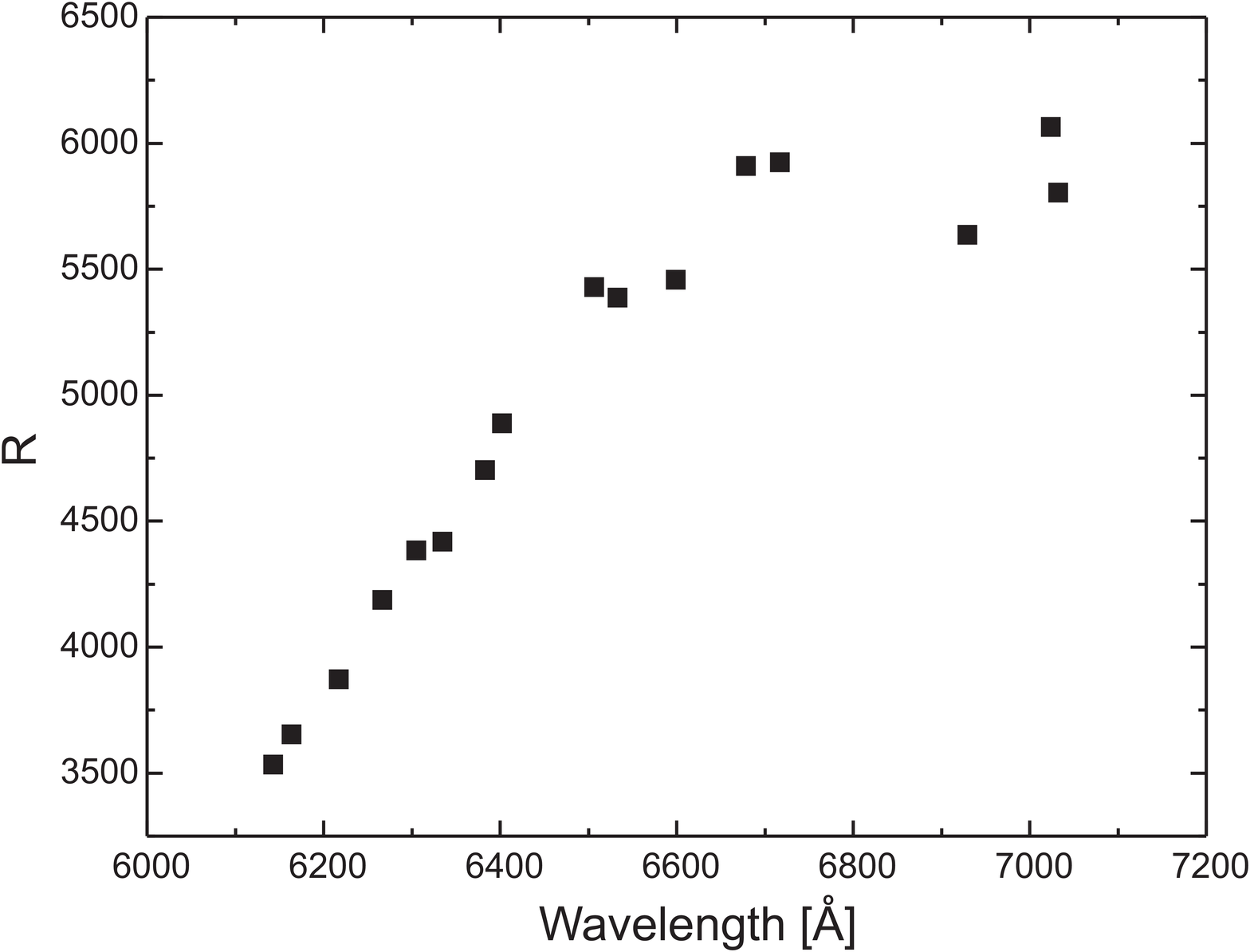}} \caption{The derived resolving power of FIASCO for the
used spectral range between 6127\,\AA~and 7061\,\AA. } \label{respow}
\end{figure}

This yields resolving powers $\lambda/\Delta\lambda$ of the spectrograph which increase from about
3500 at the blue end of the covered spectral range up to 6000 at its red end, as it is illustrated
in Fig\,\ref{respow}.

\section{First results with FIASCO at the 0.9\,m telescope}

During first light spectroscopy with FIASCO together with its calibration unit at the 0.9\,m
telescope end of may 2008 several stars with different spectral types and magnitudes were observed
to test the performance of the spectrograph during night time operation. Furthermore, the quality
of calibration data, taken with the calibration unit in the light pass, and their use for data
reduction was investigated.

The telescope acquisition of a target on the fibre entrance, the optimized telescope focusing, as
well as the spectroscopy of targets with FIASCO were all fully successful. The obtained arc and
flatfield spectra allowed an optimal calibration of the spectra.

In the following subsections we present first results, which were obtained with FIASCO while
testing the spectrograph during night time operation. All spectra were wavelength calibrated and
flatfielded with calibration spectra, taken with the calibration unit either before or after the
spectroscopy of the different targets. For flux calibration, spectra of standard stars with known
spectral types were always taken at the same airmasses as the observed targets.

\subsection{FIASCO Detection Limit}

In order to derive the detection limit of the spectrograph at the 0.9\,m telescope several spectra
of G2V stars were obtained during different observing nights but always under photometric
conditions. The maximal achieved signal-to-noise ratio of each spectrum was measured which finally
yielded the detection limit. For 600\,s of integration time a signal-to-noise ratio of $S/N=10$ can
be expected for FIASCO spectra of solar like stars with a magnitude of $V=12.1\pm0.3$\,mag.

\subsection{Lithium detection}

\begin{figure}[h!]
\resizebox{\hsize}{!}{\includegraphics[]{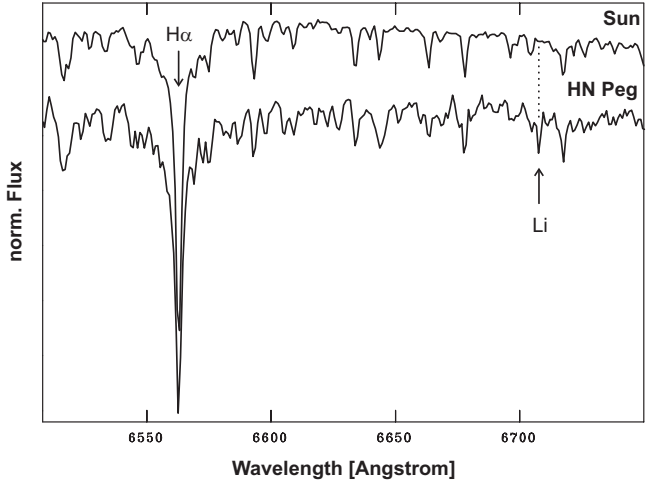}} \caption{The FIASCO spectrum of the young G0V
flare star HN Peg, taken on September 13th 2008. Two FIASCO spectra each with an integration time
of 600\,s were averaged. The continuum of the averaged spectrum is normalized. For comparison a
normalized FIASCO spectrum of the sun is shown. The Lithium absorption feature at 6708\,\AA~is
clearly detected in the spectrum of HN Peg.} \label{hnpeg}
\end{figure}

On September 13th 2008 we took two spectra with 600\,s integration time of the young flare star HN
Peg ($V=6.0$\,mag). The average of both spectra is shown in Fig.\,\ref{hnpeg}. The continuum of the
average spectrum is normalized and a normalized solar spectrum is shown for comparison. Beside the
strong H$\alpha$ absorption line at 6563\,\AA~several faint atomic absorption features are detected
in the spectrum of HN Peg, which are also found in the solar comparison spectra. The atmosphere of
HN Peg still contains Lithium, whose absorption feature is clearly detected at 6708\,\AA~in our
spectrum. In contrast Lithium is not found in the solar spectrum, as it is expected for a several
Gyr old main sequence star.

\subsection{Simultaneous Photometry and Spectroscopy}

\begin{figure}[h!]
\resizebox{\hsize}{!}{\includegraphics[]{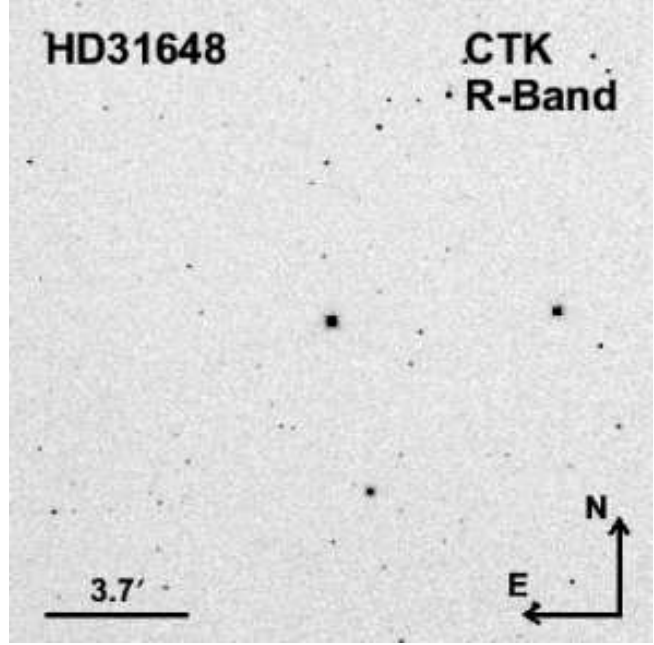}} \caption{This pattern shows a detail of
one CTK R-band image with 4\,s of integration time, which was taken during the simultaneous
spectro- photometric monitoring of the Herbig Ae star HD\,31648. The star is located in the center
of the image.} \label{hd31648ctk}
\end{figure}

In order to test the simultaneous photometric and spectroscopic monitoring of a star with the
telescopes of the University Observatory Jena on October 13th 2008 we observed the Herbig Ae star
HD\,31648 ($V=7.7$\,mag) over a span of time of 104\,min. The star was imaged with the camera CTK
(see \cite{mugrauer2009} for details) and 178 images each with an integration time of 4\,s could be
taken in R-band. The dark current and bias level of all CTK images were removed by the subtraction
of dark frames taken with the same integration time. We used the median of 5 dark frames to remove
cosmics detected in the individual dark frames. The dark and bias subtracted CTK images were then
flatfielded with sky flats taken in twilight. A detail of one of these CTK images is shown in
Fig.\,\ref{hd31648ctk}.

\begin{figure}[h!]
\resizebox{\hsize}{!}{\includegraphics[]{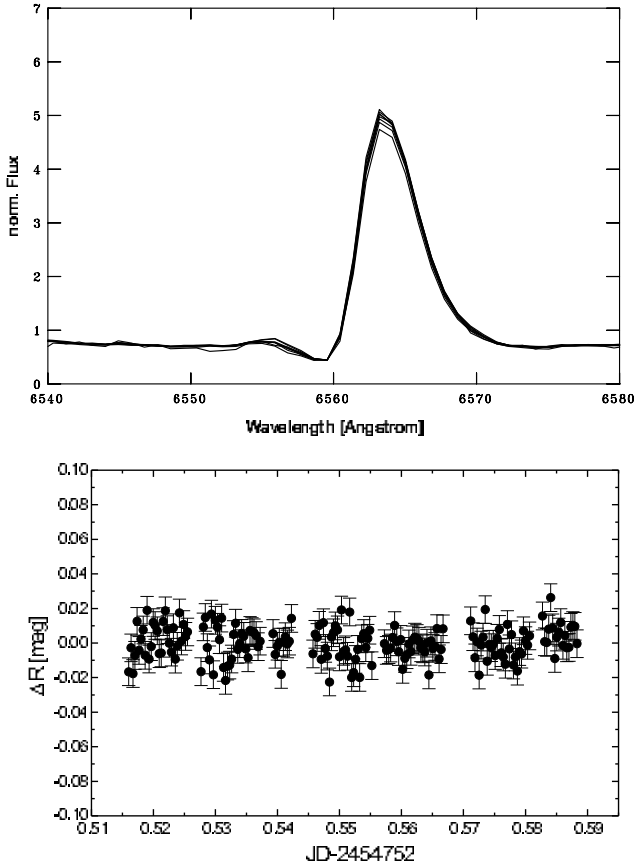}} \caption{The results of simultaneous
photo- and spectroscopy taken with CTK and FIASCO on October 13th 2008. The Herbig Ae star
HD\,31648 was observed over a span of time of 104\,min. \textbf{Top pattern:} During the
simultaneous photo- and spectroscopic monitoring of the star 8 spectra each with an integration of
600\,s were taken with FIASCO. The continua of all spectra are normalized to unity at 6600\,\AA.
The individual spectra show all the P-Cyg profile of the H$\alpha$-line. The 8 spectra are all
overplotted to each other to show their slight variability. \textbf{Bottom pattern:} Simultaneously
to the spectroscopy, 178 R-band images could be taken with the CTK, each with an integration time
of 4\,s. The photometry of the star in the given time interval is found to be stable, with a
photometric scatter relative to the average brightness of only $\pm$9\,mmag.} \label{photospec}
\end{figure}

Simultaneously to the CTK imaging 8 spectra of the star could be taken with FIASCO each with an
integration time of 600\,s. The individual spectra were normalized to unity at a wavelength of
6600\,\AA. The CTK relative photometry, as well as the simultaneously taken spectra of the star are
shown in Fig.\,\ref{photospec}. The photometry of HD\,31648 does not show a significant photometric
variability within the monitored span of time. The scatter of the CTK R-band photometry of
HD\,31648 is only $\Delta R=\pm0.009$\,mag. Fig.\,\ref{photospec} shows a display of the taken
spectra for a spectral range centered around H$\alpha$ from 6540\,\AA~to 6580\,\AA. The individual
spectra are overplotted. The P-Cyg profile of the H$\alpha$-line of HD\,31648 is clearly detected
in all spectra with a slight variation during the monitored span of time.

\subsection{Planet spectroscopy with FIASCO}

\begin{figure}[h!]
\resizebox{\hsize}{!}{\includegraphics[]{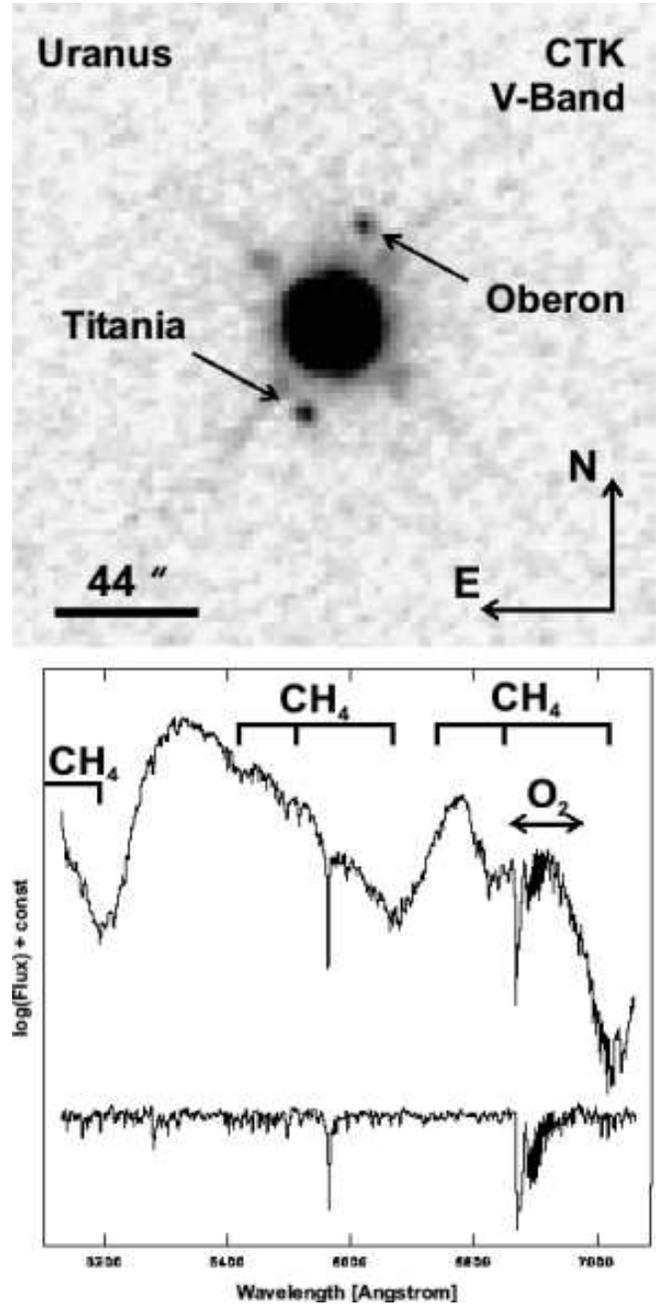}} \caption{\textbf{Top pattern:} The planet
Uranus imaged with the CTK on Sep 4th 2008. In V-Band 12 CTK images each with an integration time
of 5\,s are averaged. Beside the planet also its two moons Titania and Oberon are detected in the
average of our CTK images. \textbf{Bottom pattern:} The FIASCO spectrum of planet Uranus, taken on
Sep 4th 2008. The spectrum is the average of two 600\,s spectra of the planet. A normalized
spectrum of the sun is shown for comparison. Beside the absorption features of the reflected sun
light the Uranus spectrum also exhibits broad absorption features of Methane similar to those found
in the spectrum of Neptune's atmosphere (see Fig.\,\ref{neptun}).} \label{uranus}
\end{figure}

In September 2008 we took spectra of the outer gas giant planets Uranus and Neptune, which were
also imaged with the CTK in V-band. On Sep 4th 2008 12 images of Uranus ($V=5.8$\,mag) each with an
integration time of 5\,s were averaged to the image shown in Fig.\,\ref{uranus}. The two Uranus
moons Titania and Oberon are detected in the CTK images close to the planet. All CTK images were
reduced in the same way as described in the previous subsection.

In addition to our CTK imaging we obtained two spectra with FIASCO each with an integration time of
600\,s whose average is shown in Fig.\,\ref{uranus}. Beside the atomic absorption lines of the
reflected sun light (see normalized solar spectrum for comparison) the Uranus spectrum also shows
the broaden absorption features of Methane, a significant component of the planet atmosphere.

\begin{figure}[h!]
\resizebox{\hsize}{!}{\includegraphics[]{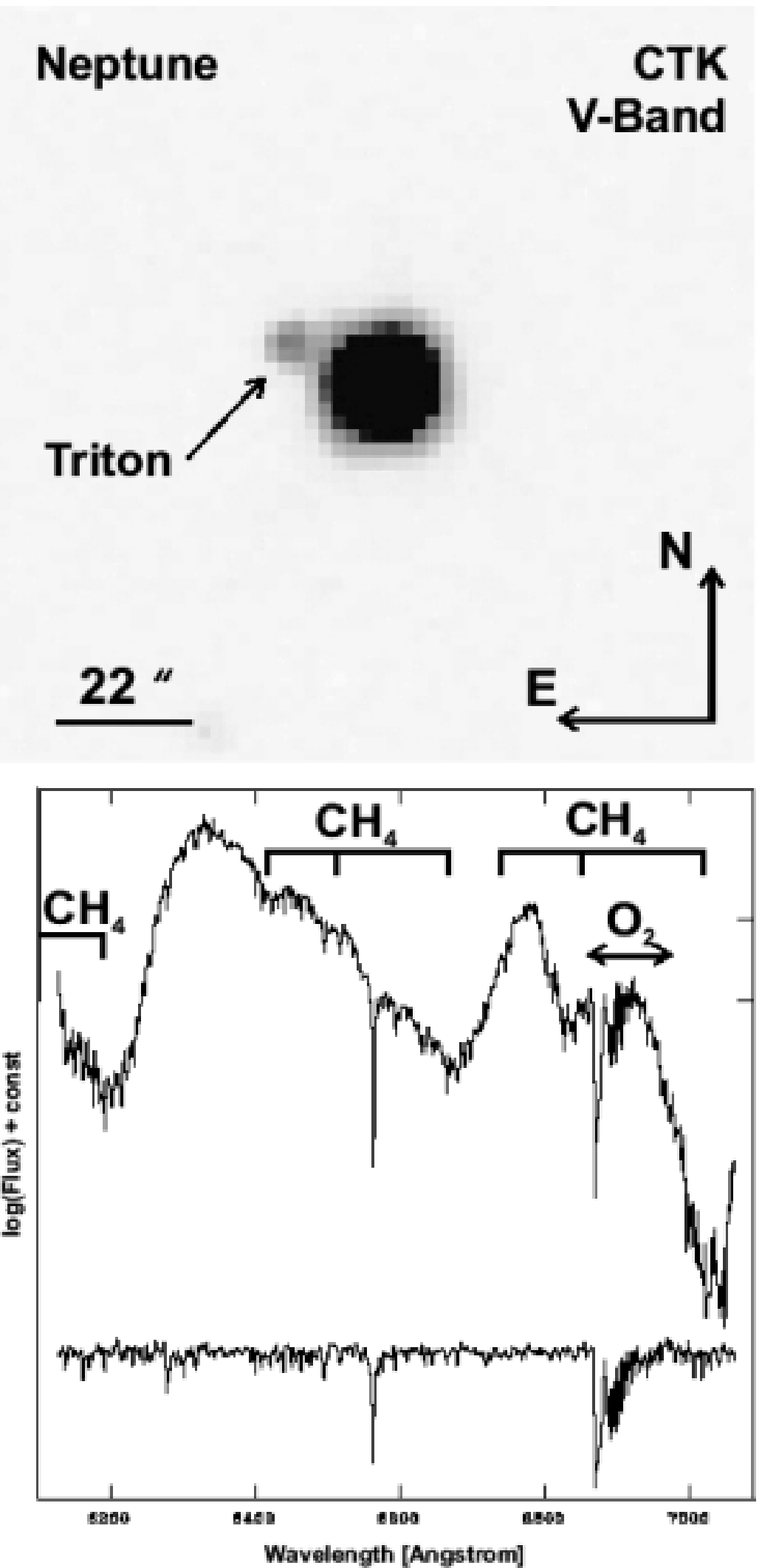}} \caption{\textbf{Top pattern:} The planet
Neptune and its moon Triton imaged with the CTK on Sep 4th 2008 in V-Band. 5 images each with an
integration time of 20\,s were averaged. \textbf{Bottom pattern:} The FIASCO spectrum of the planet
Neptune taken on Sep 18th 2008. Three spectra each with an integration time of 600\,s are averaged.
The normalized FIASCO spectrum of the sun is shown for comparison. The Neptune spectrum shows all
absorption lines of the solar spectrum due to reflected sun light on the planet's atmosphere. As in
the spectrum of Uranus (see Fig.\,\ref{uranus}) also the continuum of the Neptune spectrum is
strongly affected by broad absorption features of Methane.} \label{neptun}
\end{figure}

In the same night we also observed Neptune ($V=7.8$\,mag) with the CTK in V-band and 5 images each
with an integration time of 20\,s were taken. All CTK images are averaged to the image which is
shown in Fig.\,\ref{neptun}. This CTK image also shows the Neptune moon Triton, which is detected
close to the planet. The CTK images were reduced in the same way as described in the previous
subsection.

The spectroscopy of Neptune was not possible in the same night because the planet was already to
close to the horizon. However, on Sep 18th 2008 we could obtain first spectra of Neptune. We took
three spectra each with an integration time of 600\,s which were then averaged to the spectrum
shown in Fig.\,\ref{neptun}. As the Uranus spectrum also the Neptune spectrum shows all absorption
lines of the reflected sun light but its continuum is strongly affected by the broad absorption
features of Methane.

\subsection{Spectroscopy of faint point like objects}

\begin{figure}[h!]
\resizebox{\hsize}{!}{\includegraphics[]{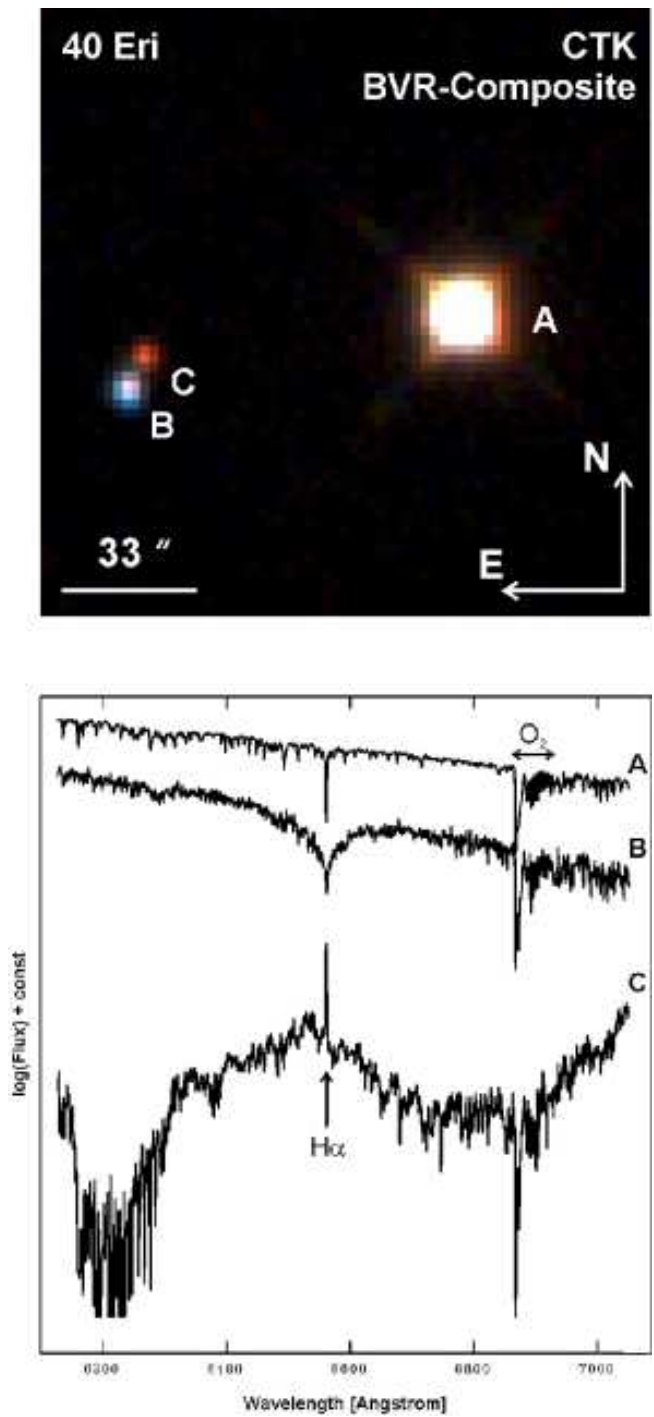}} \caption{\textbf{Top pattern:} The CTK
BVR-composite of the triple star system~40\,Eri\,ABC, taken on October 11th 2008. The integration
time in the V-, and R-band is only 1\,s, but 5\,s in B-band. \textbf{Bottom pattern:} The FIASCO
spectra of all components of the 40\,Eri system. Three spectra were taken and averaged for each
star. The individual integration times of the spectra are 300\,s in the case of 40\,Eri\,A, and
600\,s for B and C, respectively.} \label{40eriabc}
\end{figure}

In the first weeks of operation spectra of point like sources brighter than $V=10$\,mag were taken
with FIASCO which were easily visible with the fibre viewing camera. In order to test the
performance of this camera also with fainter targets we observed the triple star system 40
Eri\,ABC. This system is a hierarchical triple star composed of a K dwarf primary ($V=4.4$\,mag)
and a close pair of a white ($V=9.5$\,mag) and a M dwarf ($V=11.2$\,mag). On October 11th 2008 we
observed the system with the CTK and obtained a BVR-composite image which is shown in
Fig.\,\ref{40eriabc}. The integration time in V, and R is only 1\,s while it is 5\,s in B-band. The
separation between 40\,Eri\,A and B is about 79\,arcsec while 40\,Eri\,B and C are separated by
only $\sim$7\,arcsec, as measured in our CTK images. All CTK images are reduced in the same way as
it was already described in the previous subsections. The different effective temperatures of the
three components in the 40\,Eri system are already revealed by their photometry. The orange shining
primary 40\,Eri\,A is an K1 main sequence star, while the faint and blue B component is a DA white
dwarf. This component has a close red shining stellar companion 40\,Eri\,C which is an M4 to 5
dwarf.

We obtained spectra of all components of the 40\,Eri system which are shown in
Fig.\,\ref{40eriabc}. Telescope acquisition, in particular on the faint B and C components was
possible using the fibre viewing camera in its periodical frame read out mode with integration
times of several seconds. Although the separation between both components is only 7\,arcsec, i.e.
only slightly larger than twice the fibre diameter (2.7\,arcsec) separate spectroscopy of both
components was doable. For each component always 3 spectra were taken with individual integration
times of 300\,s in the case of 40\,Eri\,A, and 600\,s for B and C, respectively. The different
shapes of the H$\alpha$-line in the spectra of the three stars is clearly detected. While
40\,Eri\,A shows a narrow H$\alpha$-line the one in the spectrum of 40\,Eri\,B is strongly
broadened due to its high surface gravity, typical for a white dwarf. In contrast 40\,Eri\,C
exhibits H$\alpha$ in emission which indicated ongoing strong chromospherical activity.

\subsection{Spectroscopy of faint extended objects}

\begin{figure}[h!] \resizebox{\hsize}{!}{\includegraphics[]{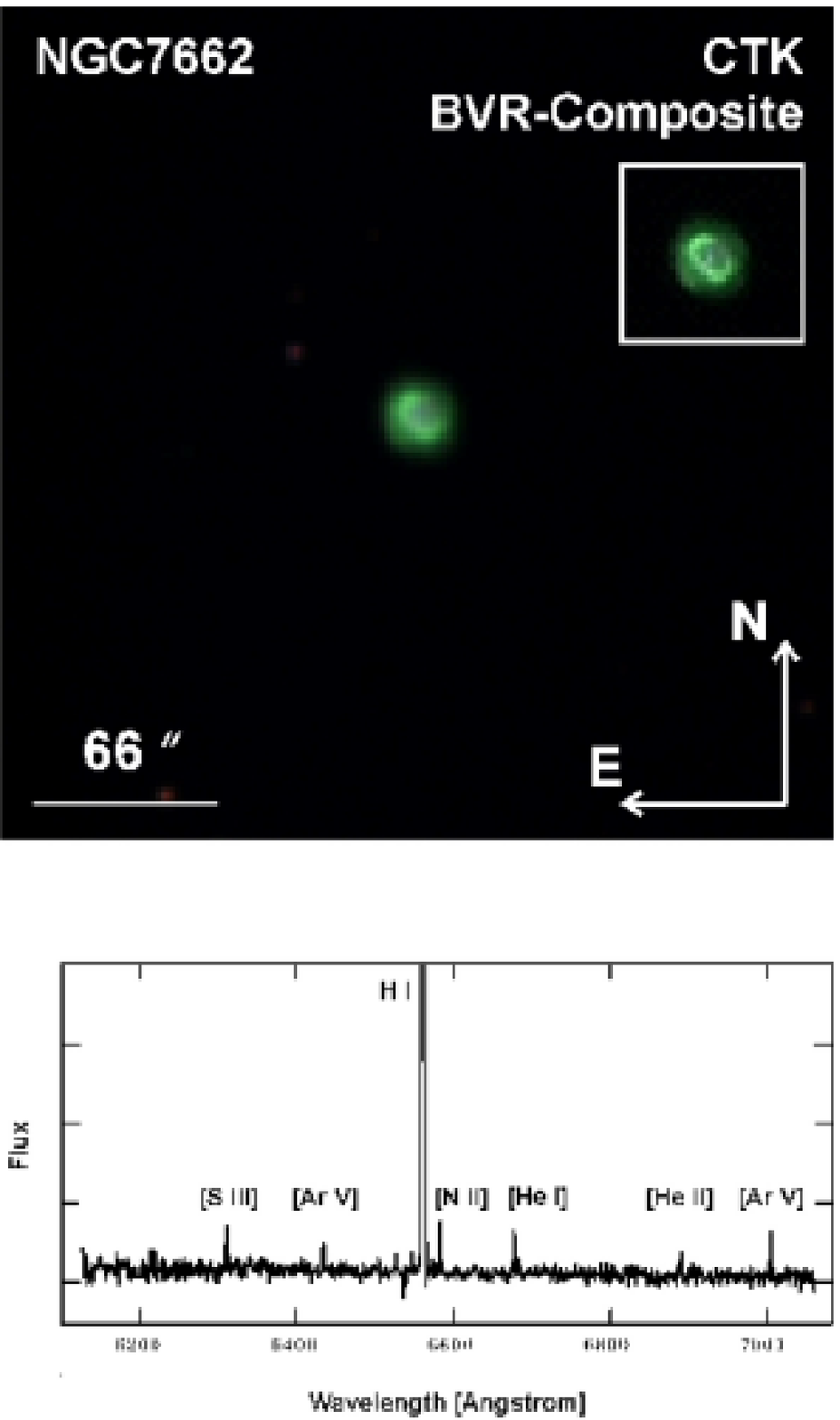}} \caption{\textbf{Top pattern:}
The CTK BVR-composite of the planetary nebula NGC7662 in the constellation Andromeda, taken on
October 11th 2008. In each filter three images were taken with individual integration times of
40\,s in B-, and 20\,s in V- and R-band, respectively. The individual CTK images of NGC7662 were
also deconvolved using surrounding stars as PSF reference (see small image in the white box).
\textbf{Bottom pattern:} The FIASCO spectrum of NGC7662 taken on October 11th 2008. It is the
average of two 600\,s spectra. Beside the strong H$\alpha$ emission line at 6563\,\AA, the
forbidden lines S [III] at 6312\,\AA, [Ar V] at 6435\,\AA~and 7005\,\AA, [N II] at 6583\,\AA, [He
I] at 6678\,\AA, and [He II] at 6891\,\AA~are detected.} \label{blau}
\end{figure}

\begin{figure}[h!]
\resizebox{\hsize}{!}{\includegraphics[]{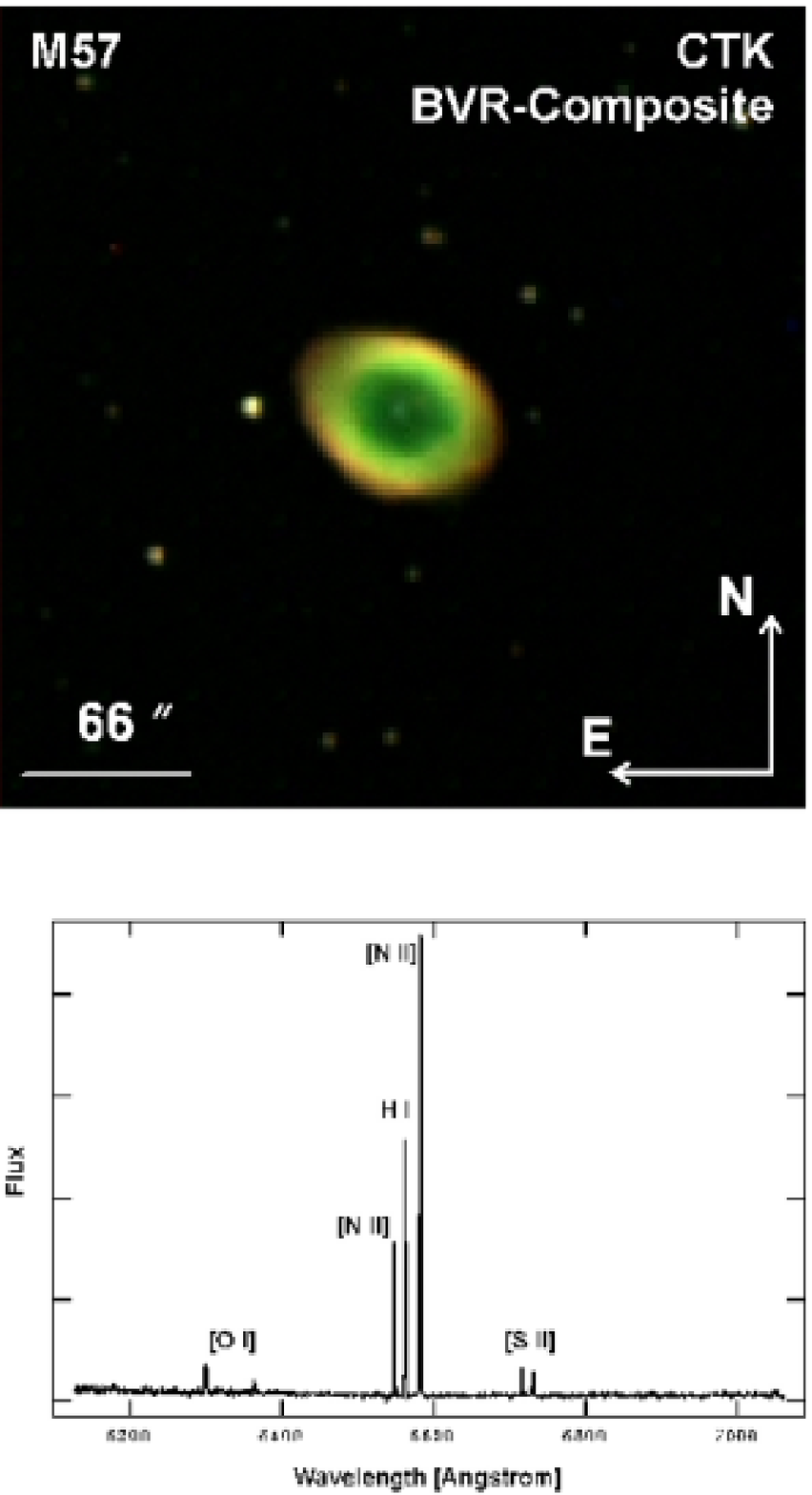}} \caption{\textbf{Top pattern:} A CTK
BVR-composite of the planetary nebula M57 in the constellation Lyra, taken on September 17th 2006.
Three CTK images each with 120\,s integration time in B-, and two in V-, and R-band, respectively,
were taken and averaged. \textbf{Bottom pattern:} The FIASCO spectrum of M57 taken on October 11th
2008. The spectra is the average of two spectra each with an integration time of 600\,s. Beside the
H$\alpha$-line at 6563\,\AA~also the forbidden lines of [O I] at 6300\,\AA~and 6364\,\AA, [N II] at
6548\,\AA~and 6583\,\AA, and [S II] at 6717\,\AA~and 6731\,\AA~are detected.\vspace{6cm}}
\label{m57}
\end{figure}

In order to test the telescope acquisition with the fibre viewing camera also on faint extended
objects we tried to take spectra of deep sky objects. Thereby, the fibre viewing camera was used in
its endless read out mode with freely adjustable integration times. On October 11th 2008 spectra of
the planetary nebula NGC7662 (Blue Snowball) in the constellation Andromeda, and the famous ring
nebula M57 (NGC6720) in the constellation Lyra were taken with FIASCO. The fibre (2.7\,arcsec of
diameter) was center in the central region of NGC7662 while its was placed on the northern arc of
M57. For each nebula two 600\,s spectra were averaged which are shown in Fig.\,\ref{blau} \&
\ref{m57}.

The most prominent feature in the spectrum of NGC7662 is the strong H$\alpha$ emission line at
6563\,\AA. In addition we could identify in the FIASCO spectrum also the forbidden lines of S [III]
at 6312\,\AA, [Ar V] at 6435\,\AA~and 7005\,\AA, [N II] at 6583\,\AA, [He I] at 6678\,\AA~and [He
II] at 6891\,\AA. In contrast, the most prominent feature in the spectrum of the northern arc of
M57 is the forbidden line of [N II] at 6583\,\AA~folowed by the H$\alpha$ emission line at
6563\,\AA~and the forbidden line of [N II] at 6548\,\AA. We could also identify the weaker features
of the forbidden lines of [O I] at 6300\,\AA~and 6364\,\AA, as well as [S II] at 6717\,\AA~and
6731\,\AA.

In addition to the spectroscopy both planetary nebulas were also imaged with the CTK. The BVR color
composite images of both nebulas are shown in Fig.\,\ref{blau} \& \ref{m57}. For NGC7662 three CTK
images were taken in each filter, each with an integration times of 40\,s in B-, but 20\,s in V-
and R-band. All CTK images of the planetary nebula were also deconvolved using surrounding stars as
PSF reference. The deconvolved BVR composite image is shown in a white box in Fig.\,\ref{blau}. The
famous ring nebula M57 was already imaged with the CTK on September 17th 2006. Images with 120\,s
of integration time were taken in each filter, three in B-, and two in V- and R-band, respectively.
The data reduction of all CTK images was done, as already described in the previous
subsections.

\acknowledgements{We would like to thank Jes\'us Rodriguez, Carlos Guirao from ESO Garching, M.
Sterzik from ESO Chile, V. Burwitz from MPE Garching, J. Alcala from INAF Capodimonte, and A.
Seifahrt from University G\"{o}ttingen for all their help. We would like also to thank the MPE for its
contribution to FIASCO hardware, as well as the staff of the mechanic and electronic division of
the faculty for physics and astronomy at the University Jena. This publication makes use of data
products from the SIMBAD and VIZIER databases, operated at CDS, Strasbourg, France.\newpage}

\newpage

\begin{thebibliography}{}
\bibitem[Avila et al.~1999]{avila1999} Avila, G., Guirao, C.,
Rodr{\'{\i}}guez, J., Alcal{\'a}, J.~M., \& Pittichov{\'a}, J.\ 1999, IAU Colloq.~173: Evolution
and Source Regions of Asteroids and Comets, 235
\bibitem[Mugrauer~2009]{mugrauer2009} Mugrauer, M.\ 2009, AN, arXiv:0903.4116
\bibitem[Pfau~1984]{pfau1984} Pfau, W.\ 1984, Jenaer Rundsch.,
29.~Jahrg., Heft 3, p.~121 - 122, 29, 121

\end{thebibliography}
\end{document}